# A 300 kA Pulsed Power Supply for LBNF

C.C. Jensen, T. Omark, H. Pfeffer, K. Roon
Fermi National Accelerator Laboratory
Batavia IL

J. Hugyik, B. Morris
SLAC National Accelerator Laboratory
Menlo Park, CA

*Abstract*— The Long Baseline Neutrino Facility (LBNF) will produce the world's most intense neutrino beam. Three series connected magnetic horns will require 5kV, 300kA, 800μs pulses at a rate of 0.7Hz to focus the beam. Fermilab has designed and built pulsed high current supplies for horns in the past. Pulsed currents of 205 kA for Neutrinos at Main Injector (NuMI / NOvA), focusing a 120 GeV beam, and 170 kA for Booster Neutrino Beam (BNB / MiniBooNE) for focusing an 8 GeV beam have been operational for about 18 years. While the magnetic horns are expected to be replaced, the LBNF horn power supply is expected to last the lifetime of the project, 30 years.

A resonant, half sine wave pulser was used for NuMI and BNB and has many practical advantages. The system has impedance limited fault currents by design, albeit large ones. Many fault modes of the system have been calculated. Several circuit changes were incorporated so that a single fault can be tolerated. The supply must also be reversible to enable both neutrinos and anti-neutrinos to be produced.

## I. Introduction

High current pulsed supplies have been built for many experiments, most recently at J-PARC for the K2K & T2K experiments [1] and at Fermilab for NuMI [2] and BNB beamlines which have also supported multiple experiments. T2K has 3 horns focusing 4.5 us beam pulse with 250 kA, ~ 2.4 ms pulse width, 0.5 pps and < 1 kV which is quite similar to NuMI with 2 horns focusing an 11 us beam pulse at 200 kA, 2.6 ms pulse width, 0.7 pps and < 1 kV. The BNB system has a single horn focusing a 1.6 us beam pulse at 170 kA, 140 us pulse width, 5 pps and < 6 kV. These have provided many lessons learned. They all use thyristor switches, capacitors for stored energy with a resonant L-R-C discharge and a transmission line or lines to deliver the high current to the horns. The horns are in a very high radiation environment. The transmission lines are significant fraction of the load inductance and resistance but the power supplies require transmission lines on the order of 30 m, to keep the electronics in a lower radiation environment.

One of the first design choices is driven by the maximum voltage applied to the transmission line and magnetic horn. This limit determines if a pulse transformer is preferred and if horns can be put in series. This transformer needs to function in the relatively high radiation area near the horns to make the voltage reduction at the horn the most effective. The choice to drive multiple horns in series is also based on allowable voltage. The second choice is the segmentation of the power supply into multiple units. This is driven by semiconductor limits on voltage and current and by reliability. The last choice is whether to recover the energy left in the capacitor after the pulse discharge. Recovering energy can reduce the size of the charging supply if losses in the transmission lines and horns are low.

The first choice for LBNF was to drive all the horns in series without a transformer. Neither NuMI nor BNB have pulse transformers. The BNB horn and transmission line operates at up to 6 kV [3] in air and the horns could be designed to operate at up to 5 kV. The LBNF horns will operate in an argon environment while most of the stripline with be in air. The use of a single supply is also a weak direct cost benefit. The stored energy is the same no matter how many supplies are used. The semiconductor cost is likely lower with a single supply, but the main power supply advantage is the reduced assembly and controls cost of a single supply. The second choice for LBNF was to have many parallel pulsers. This was done in NuMI (12 cells) and BNB (16 cells). The third choice for LBNF is to use energy recovery. Detailed simulations show only 30% of the initial stored energy is lost per pulse. NuMI and T2K do not use energy recovery because the lost energy is a large fraction of the initial stored energy while BNB does use energy recovery. The final choice is to use the odd numbered / balanced stripline for our transmission line [4]. This construction balances forces and reduces losses and inductance over an even numbered stripline.

Another consideration for LBNF is the long-term development and testing of horns. A lesson learned from the experience with NuMI and BNB is that horns require ongoing improvements. This could be based on a failure mechanism that was unforeseen or on a desire to increase performance. LBNF expects an increase in beam power as part of the program and anticipates horn improvements as well. Testing of new horns at full current, operating pulse length and repetition rate is needed to verify performance and find early failures. Various scenarios were considered, and the decision was made to build two complete horn power supplies, one dedicated to testing and one for operation. The design of the horn supply also takes this into consideration using the same detailed design for both supplies and changing only the capacitance value and charging voltage.

## II. Design and Implementation Choices

First, there are different performance limits for all the components in a pulsed supply: capacitors, inductors, semiconductor and connections. Practical considerations determined the width of the stripline and that determined losses. Having an extremely uniform magnetic field in the horn drove a requirement for currents matched better than +/-1% in each of the four striplines. The number of cells in the pulser is required to be divisible by 4 to feed the 9 layer stripline (5 supply and 4 return layers). The balanced horn current and the balanced stripline current meant the number of cells also had to be divisible by 8 so the 2 outer layers could be driven with half of the current of the inner 7 layers. This limited our choices to 8, 16 or 32. One of our internal design guides for long lifetime is to limit the thermal cycling of semiconductors during each pulse. Experience with long life cycle design to not us special purpose semiconductors that result in replacement issues in the future. The capacitor bank value and voltage depend on meeting the peak current and pulse length. Field simulations of the stripline and calculations of the horn inductances determined the total load of 3.14 µH and resistance of 0.95 mOhms. The pulse length requirement determined a total capacitance of 19 mF and the 300 kA peak current determines the nominal charge voltage of 4.3 kV. We determined that for our operating parameters, design constraints, and component constraints the choice was 32 parallel pulsers, or cells, using standard inverter grade thyristors and a maximum operating voltage of 4.7 kV. The schematic of a single cell is shown in Fig 1.

The mechanical layout and connection scheme was the next design choice. This impacts electrical performance, electromagnetic coupling, thermal performance and serviceability. The components losses were designed to not require water cooling. Arranging the highest power dissipating elements near the top will draw cooler air up and over other components. The highest losses are in the recovery choke. The saturating choke is heavy, low loss and so is placed at the bottom. Circuit topology then determines that the switch be near the bottom and above the saturating choke and the recovery diode be near the top below the recovery choke. The spacing between magnetic components is another concern because of magnetic fields. The diode and switch are mounted a coil length away to reduce magnetic field coupling. The recovery choke is air core and there are significant magnetic fields outside the saturating choke as well. The cells are arranged in pairs in a single cabinet, with coil winding orientation alternating to reduce magnetic fields away from choke as shown in Fig 3.

One of the requirements is to easily reverse the polarity of the supply. The first horn must always remain at the lowest voltage because the target is grounded and is inserted into the first horn. This means reversing the polarity of all the semiconductors. The thyristor and diode assemblies are constructed for easy reversal. A custom Litz constructed wire with 80 strands of AWG #20 with high temperature magnet wire coating is used. The larger strand size is to avoid repetitive stress failure due to pulsing. This is an equivalent wire size of AWG#2 that meets NEC code, will fit standard 3/0 lugs after it has been stripped and retains more flexibility for changing connections. These flexible leads are shown by a wavy line in Fig 1.

The connection from the capacitor bank to the switch is routed through the saturating choke using the same Litz wire as is used in the construction of the choke. A small stripline carries the pulse current from the switch output to the horn stripline and back to the capacitor bank, also a stripline, through the current transformer. A wire is also routed along the stripline to connect the recovery choke to the capacitor bank. The current per cell is 280 $A_{rms}$ in the capacitor stripline and 230 $A_{rms}$ in the saturable reactor and switch. All these efforts are to reduce inductance and stray magnetic field that can couple to other parts of the circuit.

### III. DESIGN AND FAULT TOLERANCE

Generally, a single fault should be recoverable from and shouldn't cause irreparable damage. This drives circuit design and topology. Short circuit at cell output, short circuit at load, short circuit in a capacitor, single cell switch no fire and self fire and thyristor or diode short are just some of the fault conditions we examined.

*A. Capacitor Fault*

We limit energy in a single bank to less than 10 kJ since we can't put series resistors or fuses in the pulse path. Most metal-can capacitors will contain this much energy without rupturing.

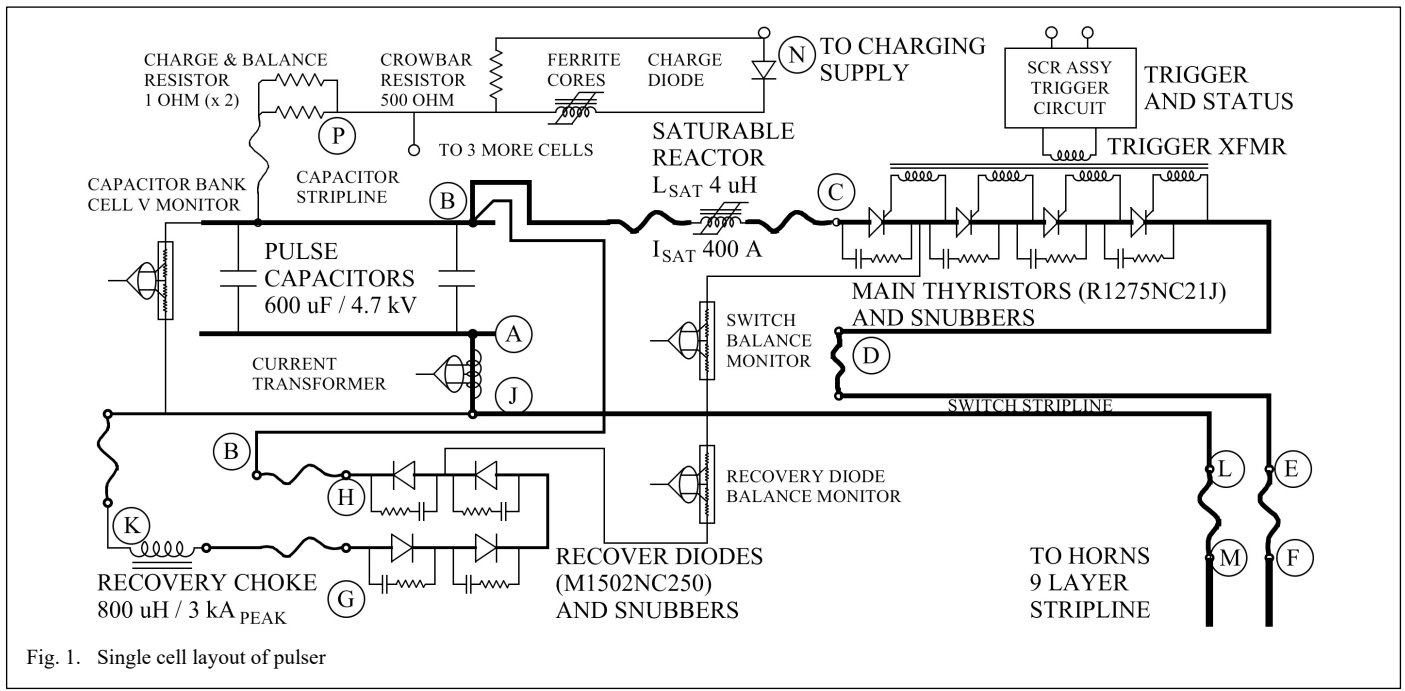

Fig. 1. Single cell layout of pulser

Experience with some self-healing capacitor technologies is that very low impedance busing can result in internal faults which do not clear. Therefore, a shorted capacitor must just be able to absorb the energy from its cell neighbors.

*B. Cell Output Fault and Saturating Choke*

If the output of a cell shorts, the inductance between the short and the cell cap bank is estimated to be 7 µH: 4 µH from the saturated choke and 3 µH from stripline and connection inductances. This will lead to a short circuit current of 30 kA$_{peak}$. A single cell firing will have a slightly higher inductance of 10 µH since it includes the total load inductance and is not the worst case, The thyristor will survive the forward current, especially since the saturating choke ensures that the thyristor is fully conducting before the high currents appear.

This short circuit will also cause a high turn off dI/dt as the thyristor current reverses. This would normally cause a high reverse current beyond which our snubber network is designed to tolerate. However the saturating choke comes out of saturation before the current reverses and reduces the -dI/dt and reverse recovery current to a manageable level. Short circuit testing during prototyping will confirm the turn off dI/dt and Qrr and may reduce the saturating choke volt second rating.

*C. Switch*

The choice of 32 cells was based on a detailed thermal and electrical model of the SCR. The forward voltage vs current was fit using (1) with A=1.23, B=–0.0588, C=0.000129 and D=0.0192 for the device chosen, R1275NC21J.

$$V_{fwd} = A + B \ln(I) + C\,I + D\,\sqrt{I} \qquad (1)$$

The energy *vs.* time into the device was simulated with a detailed thermal model of the device using Spice and common techniques [5].

With a 10 kA$_{peak}$ current, the energy per pulse in the device was 17 J and a peak junction temperature change of 11 C while with a 20 kA$_{peak}$ current the energy per pulse was 51 J with a peak temperature change of 30 C. Our design goal for long life ($10^9$ pulses) and thermal cycling at about 1 pps is 10 C. This is likely conservative, but data is sparse for long life [6] and depends upon manufacturer techniques and failure mechanisms.

A simplified estimate of average power using the forward voltage drop at full current and the average current gave 21 W while the detailed estimate above gives 24 W for a 0.7s period providing a check. Both NuMI and BNB use water cooling for their semiconductors. This has resulted in issues with non-conductive cooling lines plating up, especially for the higher voltage of BNB. We have changed all the cooling lines several times at BNB. A design for air cooling of the LBNF switch seemed feasible with this average power and would simplify maintenance considerably. The design was completed for passive air cooling while keeping electrical partial discharge requirements.

A single switch assembly not firing while the others switches do would cause the voltage across that switch to exceed the switch rating because it's capacitor would not discharge. The charging resistor connects multiple cells together and a value of 1 Ohm is enough to return the thyristor operating margin to 60%.

The SCR switch and diode assemblies required that the polarity of the devices could be switched quickly and the devices were mounted for low inductance. These requirements meant natural passive cooling via a heat sink was only possible on one side of the each of the four devices. The mounting assembly was keyed such that it would fit in either of the two switch position in a cabinet. The designed needed to provide symmetry without polarity being accidentally swapped by a technician but also allowed easy swapping of the power leads to reverse the polarity of the assembly. The assembly design also needed to be clamped with a 22 kN force and this was accomplished using an off the shelf assembly. The final design achieved minimal inductance as well as ensuring the both the high voltage and thermal cooling requirement were met. A single thyristor schematic, of 4 in a switch, is shown in Fig 2.

*D. Energy Recovery Diode and Choke*

The energy recovery diode had a similar clamping force and voltage, less power dissipation, only an RC snubber of 0.5 µF and 40 Ohms and the same voltage balancing resistor. Most of the mechanical design was carried over resulting in a similar but simpler assembly.

The energy recovery choke requirements were based on what we had success with for BNB. The main change was a reduction in losses to reduce heat load in the cabinet. It is still the highest heat dissipating elements in each cell.

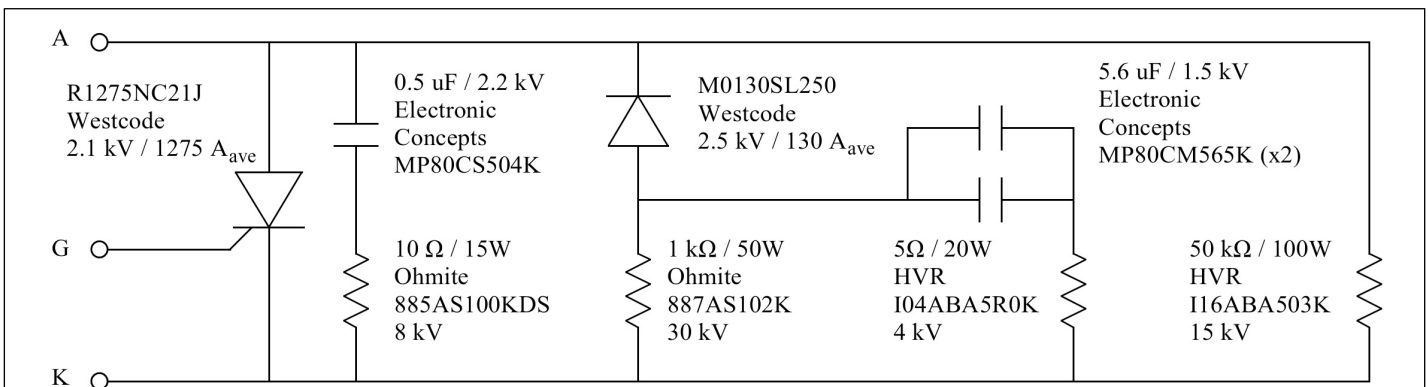

Fig. 2. Schematic of single thyristor in the four thyristor switch assembly showing part values and manufacturer

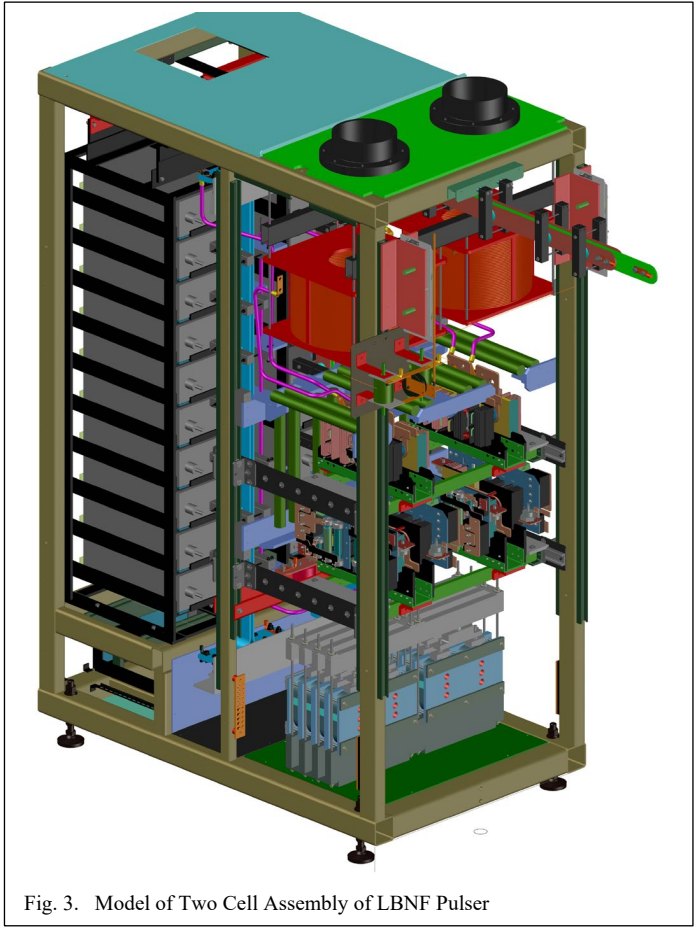

Fig. 3. Model of Two Cell Assembly of LBNF Pulser

## A. Energy Recovery Choke

We specified air-core choke of 800 µH, to operate at 5kV$_{peak}$, 3 kA$_{peak}$ with losses less than 400 W at 120 Arms at 250 Hz. We ordered chokes from two vendors. Partial discharge testing showed inception of << 100 pC at 60 Hz at 6 kV$_{rms}$ from both vendors. The inductance from vendor A was low at 710 µH and the losses, based on ESR at 250 Hz, were high at 490 W. The inductance from vendor B was 800 µH and the losses were less than 400 W again based on ESR. Two different construction techniques resulted in a physical dimensions that are quite similar but different AC performance. Pulse testing will determine which construction method will last.

## B. Saturating Choke

We designed the saturating choke based on the success and experience from our BNB design. The clamping force was improved to reduce vibration and we added other mechanical features to aid in mass production. We measured the partial discharge inception voltage at about 5 kV$_{rms}$ and extinction at about 4.2 kV$_{rms}$. This is slightly below our design goal of 6 kVrms. We measured an unsaturated inductance of 410 µH and measured the saturation current of 300 A. The saturated inductance was calculated to be 4 µH

## C. Switch

Partial discharge testing showed << 100 pC at 60 Hz at 6.5 kV rms. The clamping force was verified using Fujifilm Medium prescale paper and measured a uniform compression across each thyristor.

The maximum temperature rise of the heatsink was found via testing to be approximately 16.5 C. This gives a maximum junction temperature of 67 C with 40 C ambient and 11 C cyclic, well within our experiential design constraints of 90 C maximum. Looking back at a 16 cell solution with 3 time the temperature rise for both cyclic and average heating the choice of 32 cells was the best.

## E. Monitors

We monitor the voltage balance among the four switches and four diodes. If one shorts during operation there is enough voltage margin in the remaining devices to withstand the voltage. We are operating at ~ 50% of device rating but reserve some margin for a change in load and these conditions. The use of only three devices would provide insufficient margin for load changes or fault conditions.

We also monitor the capacitor current and voltage in each cell. All cells are compared against the average during the pulse to detect an imbalance in either current or voltage. This can be used to find poor connections or reduction in capacitance.

Finally, both the voltage and current monitors are differential reduces noise pickup and also leads to a much quicker reversal of polarity. The balanced, differential voltage monitors we designed and have used in several high power modulators while the current monitor is a custom order from Pearson.

## IV. TESTING

All the elements for a single cell test have been received except the mechanical structure and striplines. We have tested all the electronic elements as much as possible. Once the mechanical parts have arrived, pulsed testing during normal and fault conditions will be done.

ACKNOWLEDGMENT

A substantial amount of the initial simulation work was done by George Krafczyk and Ken Quinn.